\documentclass[prd,twocolumn,superscriptaddress,floatfix,amsmath,amssymb]{revtex4-1}
\usepackage[english]{babel}
\usepackage{graphicx}
\usepackage{dcolumn}
\usepackage{bm}
\usepackage{verbatim}
\usepackage{mathrsfs}
\usepackage{cancel}
\usepackage{color}
\newcommand{\ket}[1]{\left| {#1} \right\rangle}
\newcommand{\bra}[1]{\left\langle {#1} \right|}

\newcommand{\eq}[1]{(\ref{#1})}

\def\slashchar#1{\setbox0=\hbox{$#1$} 
\dimen0=\wd0 
\setbox1=\hbox{/} \dimen1=\wd1 
\ifdim\dimen0>\dimen1 
\rlap{\hbox to \dimen0{\hfil/\hfil}} 
#1 
\else 
\rlap{\hbox to \dimen1{\hfil$#1$\hfil}} 
/ 
\fi}

\begin{document}

\title{Wavepacket detection with the Unruh-DeWitt model}
%

\author{Eduardo Mart\'{i}n-Mart\'{i}nez}
\affiliation{Institute for Quantum Computing, Department of Physics and Astronomy and Department of Applied Mathematics, University of Waterloo, 200 University
Avenue W, Waterloo, Ontario, N2L 3G1, Canada}
\affiliation{Instituto de F\'{i}sica Fundamental, CSIC, Serrano 113-B, 28006 Madrid, Spain}
\author{Miguel Montero}
\affiliation{Instituto de F\'{i}sica Fundamental, CSIC, Serrano 113-B, 28006 Madrid, Spain}
\author{Marco del Rey}
\affiliation{Instituto de F\'{i}sica Fundamental, CSIC, Serrano 113-B, 28006 Madrid, Spain}

\begin{abstract}
In this paper we deal with several issues regarding the localisation properties of the Unruh-DeWitt (UdW) detector model. Since its original formulation as a pointlike detector, the UdW model has been used to study extensively the physics of quantum fields in presence of accelerations or curved backgrounds. Natural extensions of it have tried to take into account the spatial profile of such detectors, but all of them have met a series of problems in their spectral response which render them useless to study some of the most interesting physical scenarios. In this paper we provide a derivation of the smeared UdW interaction from QED first principles, then we analyze the spectral response of spatially smeared UdW detectors, and discuss the kind of spatial profiles which are useful for the study of relevant cases.\end{abstract}

\maketitle

\section{Introduction}\label{sec1}

The Unruh-DeWitt (UdW) model describes phenomenologically a monopole detector coupled to a massless scalar field, moving in the four-dimensional Minkowski space.  Since its inception, it has been used to study the response of detectors experiencing acceleration, to provide a proof for the Unruh effect, and particularly as  one of the main tools to probe dynamics of entanglement in the context of the recent field of Relativistic Quantum Information (RQI).

Usually, the detector considered is a quantum system with two internal states, ground state $|g\rangle$ and excited state $|e\rangle$, with $\Omega$ (taking $\hbar=1$) being the energy difference between the two levels. The detector is then coupled to the real massless scalar field $\phi$ according to the following interaction Hamiltonian:
\begin{equation}
H_{\text{int}}= \lambda \ \xi(\tau) \mu(\tau) \phi(\bm x(\tau))\label{udw0}
\end{equation}
where $\lambda$ is the coupling strength, $\xi$ is a switching function which activates during the interaction time , $\mu(\tau)$ the monopole momentum operator and $x(\tau)$ the worldline of the atom.

In spite of the differences between this monopole-scalar field interaction and QED (for instance in the behaviour at very extreme frequencies which may quantitatively vary), it characterises adequately the matter-radiation interaction in some specific settings \cite{ScullyBook} (see section \ref{sec2} for further details), while it very accurately models the interaction of internal degrees of freedom of atoms with phonon fields (for example the spin-phonon interaction of ions in a Coulomb crystal, collective excitations of Bose-Einstein condensates \cite{BoseExcitations} and other solid state and analog systems). This model and certain variations of it have been extensively used in the literature for many purposes \cite{Birrell}, including thermalization dynamics and decoherence (\cite{UnruhZurek, Massar} and references therein), although it is more known for what regards the studies of the Unruh effect and Hawking radiation \cite{Unruh,Casadio1,Casadio2}. 

As a detector model, it performs commonly under the pointlike approximation, i.e. it has no extension and interacts with the field only in the exact geometric point of the space-time where it is placed. While this assumption --which will always be an approximation since any physical detector has a finite size-- seems to be valid in many scenarios, it is not valid in general even for physically interesting scenarios, and is particularly problematic in some specific settings that we will discuss below. Also, it presents UV divergences as any pointlike interaction and cannot be guaranteed to hold for any context where we consider several detectors undergoing relativistic motion where the pointlike approximation may be violated from some reference frames. Moreover, additional problems with the pointlike nature of the detector arise. For instance, there are various regularisation schemes which yield different transition probabilities \cite{Schlicht}.

For all these reasons, and keeping in mind that any realistic particle detector has a finite size, it is important to model and understand particle detectors that present a spatial smearing. However, previous localisation models present a series of issues when it comes to analysing non-vacuum field states. In this paper we will show to what extent an Unruh DeWitt detector is a reliable model of electromagnetic atomic transitions, by explicitly analysing the relationship between the atomic wavefunctions and the spatial smearing. We also intend to provide a pedagogical description of the use of a spatially smeared UdW model and we will discuss how to overcome the problems when analysing signals by means of a small but essential modification of the spatial profiles employed in the past. Besides, we will focus on the particular case of spatially smeared uniformly accelerated detectors.

 The paper is organised as follows: In section \ref{sec2} we show from first principles how to relate the spatial profile of the UdW model to the wavefunctions of physical systems under standard QED interactions. In section \ref{sec3} we present the localisation issues of the canonical UdW detector employed in the literature when the size of the detector is comparable to the wavelength they are tuned to. In section \ref{sec4} we  propose a way around these difficulties by modifying the spatial profile of the smeared UdW detector. In section \ref{sec5} we discuss how to use these detectors to analyze arbitrary signals in accelerated settings. Finally, section \ref{conc} contains our conclusions.

\section{Modelling atomic physics with the Unruh DeWitt detector}\label{sec2}

An UdW detector is an ad-hoc phenomenological model commonly used to study idealised situations in field theory and non-inertial settings.  The model is built specifically for its useful properties and simplicity. While desirable traits are good guidelines for model building, one should always keep the physics in mind. This section is concerned with the build up of a smeared UdW detector out from first principles and standard QED interactions.

First, note that the simple scalar field model \eq{udw0} cannot be directly used to relate the UdW model to electromagnetic phenomena due to the vector character of the photon field. The vector version of an UdW interaction with a smeared field operator would be
\begin{align}H_\text{I}=\sum_{\lambda=+,-} \int d\bm x\ \lambda[\bm F(\bm{x})\sigma^++\bm F^*(\bm{x})\sigma^-] \cdot\bm{A}(\bm{x})\label{udw}\end{align}
where we have omitted any switching function, as the electromagnetic interaction cannot be switched, and where $\sigma^-$ is the two-level system lowering operator, as is common in the literature. We have also allowed for a complex profile function. The detector is assumed to be inertial; we discuss the treatment of an accelerated UdW detector in sec. \ref{sec5}.

The physical system the UdW detector tries to emulate is that of a two-level atom coupled to a quantum electromagnetic field. The Hamiltonian for such a system is well-known and it is simply
\begin{align} H_\text{I}^\text{QED}&=e\bm{p}_\text{D}\cdot\bm{A}(\bm{x},0)\nonumber\\&=\bm{p}_\text{D}\cdot \sum_{\lambda=+,-} \int \frac{d\bm p}{\sqrt{2p}}\left[\epsilon_{\bm{p},\lambda}a_{\bm{p},\lambda}^\dagger e^{-ipx}+\epsilon^*_{\bm{p},\lambda}a_{\bm{p},\lambda}e^{ipx}\right],\label{qedh}\end{align}
where $\bm{p}_\text{D}$ is the detector momentum and in the last  two equalities we assume a $(1+1)$-dimensional setting. In this setting, $\bf{p}_\text{D}$ is itself an operator, the momentum operator of the valence electron of the two-level system. There is a simple way to relate \eq{qedh} to \eq{udw}; we simply write down the operator in \eq{qedh} in terms of field operators and atomic Pauli matrices.  There are four possible matrix elements for the $\bm{p}_\text{D}\bm{A}(\bm{x},0)$ operator in terms of the relevant wavefunctions, $\Psi_g(\bm x)$ for the ground state and $\Psi_e(\bm x)$ for the excited state of the detector, which can be neatly written into matrix form as,
\begin{align}H_\text{I}^\text{QED}&=\alpha\mathbf{I}+\beta\sigma_z+\gamma\sigma_x+\delta \sigma_y,\nonumber\\
\alpha&=e\sum_{\lambda=+,-}\int \frac{d\bm p}{\sqrt{2p}}\left[a^\dagger_p\frac{G^\lambda_{gg}(\bm p)+G^\lambda_{ee}(\bm p)}{2}+\text{H.c.}\right],\nonumber\\
\beta&=e\sum_{\lambda=+,-}\int \frac{d\bm p}{\sqrt{2p}}\left[a^\dagger_p\frac{G^\lambda_{gg}(\bm p)-G^\lambda_{ee}(\bm p)}{2}+\text{H.c.}\right],\nonumber\\
\gamma&=e\sum_{\lambda=+,-}\int \frac{d\bm p}{\sqrt{2p}}\left[a^\dagger_p\frac{G^\lambda_{ge}(\bm p)+G^\lambda_{eg}(\bm p)}{2}+\text{H.c.}\right],\nonumber\\
\delta&=e\sum_{\lambda=+,-}\int \frac{d\bm p}{\sqrt{2p}}\left[a^\dagger_p\frac{G^\lambda_{ge}(\bm p)-G^\lambda_{eg}(\bm p)}{2i}+\text{H.c.}\right],\label{gen}\end{align}
with
\begin{align}G^\lambda_{ij}(\bm p)=\int d\bm x\ e^{-ipx} \bm{\epsilon}_{\bm{p},\lambda}\cdot(\Psi^*_i(x) [-i\bm{\nabla} \Psi_j(x)]).\label{u1}\end{align}

If we performed the same calculation with the interaction \eq{udw}, we would obtain
\begin{align}G^\lambda_{ij}(\bm p)=[\delta_{ig}\delta_{je}+\delta_{ie}\delta_{jg}]\int d\bm x\ e^{-ipx} \bm{\epsilon}_{\bm{p},\lambda}\cdot\bm{F}(\bm{x}).\label{u2}\end{align}

We have thus expressed the physical interaction hamiltonian $H_\text{I}^\text{QED}$ in the language of \eq{udw}. If we only consider the $\sigma_x$ and $\sigma_y$ terms, we may compare directly to \eq{udw}. From \eq{u1} and \eq{u2} we find that the two Hamiltonians are equivalent with a smearing function
\begin{align}\bm F(\bm x)=-i\Psi^*_e(\bm x) \bm{\nabla}\Psi_g(\bm x).\label{smear}\end{align}

We have thus made a first connection between \eq{udw} and the physics - \textit{ the smearing function can be obtained in terms of the atomic wavefunctions of the two-level system. }This means that the smeared UdW Hamiltonian commonly used in the literature can be related in a direct manner to the physical properties of the underlying system, directly relating the smearing function to the wavefunctions of the excited and ground states of the two-level atom. Note that the terms with $\mathbf{I}$ and $\sigma_z$ do not vanish and can never do so unless $\Psi_e=\Psi_g=0$, or in the dipolar approximation, where $e^{-ipx}\simeq 1$. 

The $\alpha$ term can be dealt with full generality, as it can be reabsorbed into the free field Hamiltonian $H_\text{F}$,
\begin{align}H_\text{F}+\alpha&=\int d\bm p\left[(\vert p\vert a^\dagger_{\bm p} a_{\bm p}+\frac{1}{\sqrt{2p}}\left(a^\dagger_{\bm p}\frac{G^\lambda_{gg}(\bm p)+G^\lambda_{ee}(\bm p)}{2}\right.\right.\nonumber\\&\left.\left.+a_{\bm p}\frac{G^\lambda_{gg}(\bm p)^*+G^\lambda_{ee}(\bm p)^*}{2}\right)\right]\end{align}
and so defining new modes
\begin{align}b_{\bm p}=a_{\bm p}+\frac{e}{(2p)^{3/2}}[G^\lambda_{gg}(\bm p)+G^\lambda_{ee}(\bm p)]\end{align}
and neglecting the usual infinite zero-point contribution, we deal with the $\alpha$ term. We only have to substitute the $a_{\bm p}$ in terms of the $b_{\bm p}$ in $\gamma$, which amounts to the addition of a constant term to $\gamma$,
\begin{align}\alpha_\gamma=\frac{e^2}{4}\Re\left\{\int \frac{d\bm p}{p}[G^\lambda_{gg}(\bm p)^*+G^\lambda_{ee}(\bm p)^*)(G^\lambda_{ge}(\bm p)+G^\lambda_{eg}(\bm p)]\right\}.\end{align}
This will induce an extra $\alpha_\gamma \sigma_x$ term in the Hamiltonian, which will be relevant or not depending on how $\alpha_\gamma$ compares with $\Omega$, the detector system gap. As $\alpha_\gamma/e$ is typically of order $1$ or less, this term will not be important if we are in a perturbation theory regime where the coupling $e$ is assumed to be small. The same considerations apply to $\alpha_\delta$. The analogous correction to $\beta$,
\begin{align}\alpha_\beta=\frac{e^2}{4}\Re\left\{\int \frac{d\bm p}{p}[G^\lambda_{gg}(\bm p)^*+G^\lambda_{ee}(\bm p)^*][G^\lambda_{gg}(\bm p)-G^\lambda_{ee}(\bm p)]\right\},\end{align}
can be reabsorbed into $\Omega$.

Dealing with $\beta$ is a more challenging matter. We cannot do the same as before because, even though we could make the Hamiltonian look like that of a free field plus an UdW interaction, the detector and field operators would not commute and hence, even without the interaction, the theory would not be a free theory.

There is one special circumstance in which $\beta$ vanishes: in systems with a strong spin interaction, so that the gap comes from the spin dependence of the energy levels. This could happen, for instance, in states of an atom within a strong magnetic field. In this case the atomic wavefunctions of the ground and excited states are the same and therefore $\beta=0$ exactly. The energy gap is $\hbar\Omega=\mu_\text{B}B$. The coupling constant to the electric field is $\approx ed$ where $d$ is a typical dimension of the atom, so in order to be in perturbation theory regime we would require electric fields of order $E<\mu_\text{B}B/ed$.

As a particular example, consider the smearing function for a hydrogen atom in its $1s$ state subjected to a magnetic field. According to \eq{smear}, we would have
\begin{align}\bm{F}(\bm{x})=-i\frac{e^{-r/a_0}}{\pi a_0^4}\bm{u}_r.\end{align}

\section{Localisation issues of the UdW detector}\label{sec3}

The first UdW localization model was introduced by Schlicht \cite{Schlicht} to solve the problems with the non-equivalence of regulators derived from the pointlike nature of the detector. In particular, he  proposed a localised spatial profile for the detector (which for computational convenience was chosen to be Lorentzian). This localisation model was further studied by Langlois \cite{Langlois2006} first, and then by Satz and Louko \cite{Satz1,Satz2}, who envisioned a more general scheme which allowed general spatial profiles to be considered undergoing arbitrary movement throughout spacetime. In these works the interaction Hamiltonian is defined as follows:
\begin{align}\label{Hamil1}
\nonumber H_{I}&=g\int_0^\infty \!\!  \frac{d \bm{k}}{\sqrt{2\omega(2\pi)^3}} \int d\bm x\ F(\bm x)\left(\sigma^+e^{i\Omega t}+\sigma^-e^{-i\Omega t}\right)\\
&\times \left(a^\dagger_{\bm{k}} e^{-i(\bm k\cdot \bm x -\omega t )}+a_{\bm{k}} e^{i(\bm k\cdot \bm x -\omega t )}\right)
\end{align}
Where $F(\bm x)$ is the spatial smearing of the detector that is supposed, for simplicity and without loss of generality, at rest and centred in $\bm x=0$, and $\Omega$ represents the frequency gap of the two-level system, in other words, the transition energy between the ground and excited state of the detector. The detector is supposed to be tuned to this frequency, i.e. it is more likely that the detector absorbs field quanta of this frequency than anything else, as we will discuss below.
In the case that the detector is point-like $F(\bm x)=\delta(\bm x)$, this model becomes the standard UdW detector introduced in \cite{DeWitt}.

 The form of the function $F(\bm x)$ must be related to the characteristics of the physical system modelled by the Hamiltonian \eq{Hamil1}. In the particular case of a two-level atom, $F(\bm x)$ should be obtainable from the wave functions of the ground and excited states of the atom and the matter-radiation interaction Hamiltonian. For the case of atomic spin transitions, the form of the Hamiltonian was derived from first principles in section \ref{sec2}.

However, it is interesting to be able to consider detectors whose size becomes comparable with the wavelength to which they are tuned. These regimes cover a great range of extremely interesting physical scenarios, e.g.  quantum microwave antennae (for example flux or charge qubits in cQED), Rydberg atoms and cavity based detectors \cite{Cavities, Circuits}, where one can no longer use an atomic wave-function to obtain the form of the Hamiltonian. Yet, it is well known that the point-like model is a good effective description of the physics \cite{Circuits,Cavities}. As we will discuss below, a question arises when studying the compatibility of  the standard spatially smeared UdW model with detectors whose characteristic length is comparable to the wavelength detected beyond the atomic scale.

In the following paragraphs we will point out a fundamental issue with the use of the traditional smeared UdW model when considering spatially extended detectors. For these cases, we propose a way to modify the detector model in order to formulate an effective theory reproducing the correct phenomenology.

Previous works dealing with the localised UdW model just considered the behavior of the detector interacting with the Minkowski vacuum, which is known to have equivalent behavior for all frequencies \cite{Satz1,Satz2}. In that respect, the problems of the model dealt with in this manuscript have not been studied yet.  We will discuss below how they can build up when one tries to process physical signals and photon wavepackets with such a detector.

For most recent analyses \cite{Schlicht,Langlois2006,Satz1,Satz2} a real symmetric profile function was chosen. In particular, the spatial profile used for most calculations was a Lorentzian. To illustrate here the problem in the most simple way we will consider a Gaussian profile, but all results apply equivalently to the Lorentzian case or to any other spatial profile.

From the Hamiltonian \eqref{Hamil1}, the integral over $\bm x$ takes the form of a trivial Fourier transform
\begin{align}
\nonumber H_{I}&=g\int_0^\infty \! \!     \frac{d \bm{k}}{\sqrt{2\omega_{\bm k}(2\pi)^3}}\left(\sigma^+e^{i\Omega t}+\sigma^-e^{-i\Omega t}\right)\\
&\times \left(\hat F(\bm k) a^\dagger_{\bm{k}} e^{i\omega_{\bm k} t }+\hat F(-\bm k)a_{\bm{k}} e^{-i \omega_{\bm k} t }\right)
\end{align}
where  we have made the dispersion relation explicit $\omega_{\bm k}=c|\bm k|$ and
\begin{equation}\label{foutra}
\hat F(\bm k)=\int d\bm x\ F(\bm x) e^{-i\bm k \cdot \bm x}\end{equation}
is the Fourier transform of the spatial profile.

We can rewrite the Hamiltonian in a way in which the resonant and anti-resonant terms are made explicit:
\begin{align}
\nonumber H_{I}&=g\int \frac{d \bm k}{\sqrt{2\omega_{\bm k}(2\pi)^3}}\left[  \hat F(\bm k)\left(a^\dagger_{\bm{k}} \sigma^-e^{i(\omega_{\bm k}-\Omega )t}+\text{H.c.}\right)\right.\\
&\left.+\hat F(-\bm k)\left(a^\dagger_{\bm{k}} \sigma^+e^{i(\omega_{\bm k}+\Omega )t}+\text{H.c.}\right)\right]
\end{align}

The time evolution operator is computed as the time ordered exponential  of the Hamiltonian. When integrating over times, the exponential factors in the Hamiltonian above are highly oscillating except when $\omega_k=c|\bm k|\approx\pm\Omega$ (stationary phase). This is the mathematical reason why a detector is tuned to the frequency of the energy gap between the ground and the excited state, as it is very well known from the study of the  matter-radiation interactions \cite{ScullyBook,JaynesCummings}. In plain words, if we want to stimulate the transition between ground and excited state we have to 'beam' the detector with radiation tuned to the natural frequency of the transition (on resonance). Otherwise, the probability of transition quickly decreases with the detuning between this natural frequency and the frequency of the radiation stimulating  the transition.

Here is the issue. If we choose $F(x)$ to be a localised smooth function such as a Gaussian or a Lorentzian, which is the case for most realistic atoms , the frequency profile $F(\bm k)$ will be a localised function centred in $\bm k=0$. Being this so, its evaluation at $\Omega/c$ will give a negligible value, for $\Omega$ sufficiently large.

The reason why this issue does not arise in electronic transitions for atoms at rest is because, for most cases, $\Omega$ is small enough. For instance, electronic transitions in the hydrogen atom have an $\Omega$ in the visible range of the spectrum, whereas the Fourier transform of the spatial profile has a width of $\sim a_0^{-1}$, which extends up to the X-ray spectrum.   

However, when we consider accelerated detectors, the Minkowski frequency for a packet centered in $\Omega$ as seen from the detector, varies effectively as a function of time as $\omega_R=\Omega e^{a\tau/c}$ (See derivation on section \ref{sec5} and \cite{Diegger}) and even for very small times it goes out of resonance. Even if we compensate the Doppler shift of the wavepacket tuning the detector in real time for the period while packet and detector overlap, we would easily get the problem of the frequency getting too far from our detector support function. If the spatial profile function does not have information about the energy gap between the ground and excited state of the detector, the response of the detector to the resonance frequency (the frequency which, by far, mostly contributes to the estimated transition from the ground and excited state) will be exponentially dampened by the Gaussian or Lorentzian tails. That implies that an accelerated detector would be, in practical terms, incapable of detecting a wavepacket centred on its natural frequency. If we are to analyse signals with UdW detectors, the model should be accordingly modified to avoid this issue. 

To illustrate the problem let us consider the most simple 1-D case, and a detector with a Gaussian spatial profile. We can take $F(x)$ to be a normalised Gaussian profile with characteristic length $L$:
\begin{equation}\label{gauss}
F(x)=\frac{1}{L\sqrt{2\pi}}\exp\left({\frac{-x^2}{2L^2}}\right)
\end{equation}
And so its Fourier Transform $\hat F(k)$ will be a Gaussian localised around $k=0$
\begin{equation}
\hat F(k)=\exp\left(\frac{- k^2L^2}{2}\right)
\end{equation}

Any frequencies such that $\omega_k\gg 0$ would be exponentially dampened in the integral over $k$ by the weight $\hat F(k)$. In particular, if $\Omega>>0$, the stationary phase contribution $\omega_k=\pm\Omega$ will be zero due to $F(\pm\Omega/c)\approx 0$,  effectively cancelling any non-trivial time evolution.

\begin{figure}[h]
\includegraphics[width=0.45\textwidth]{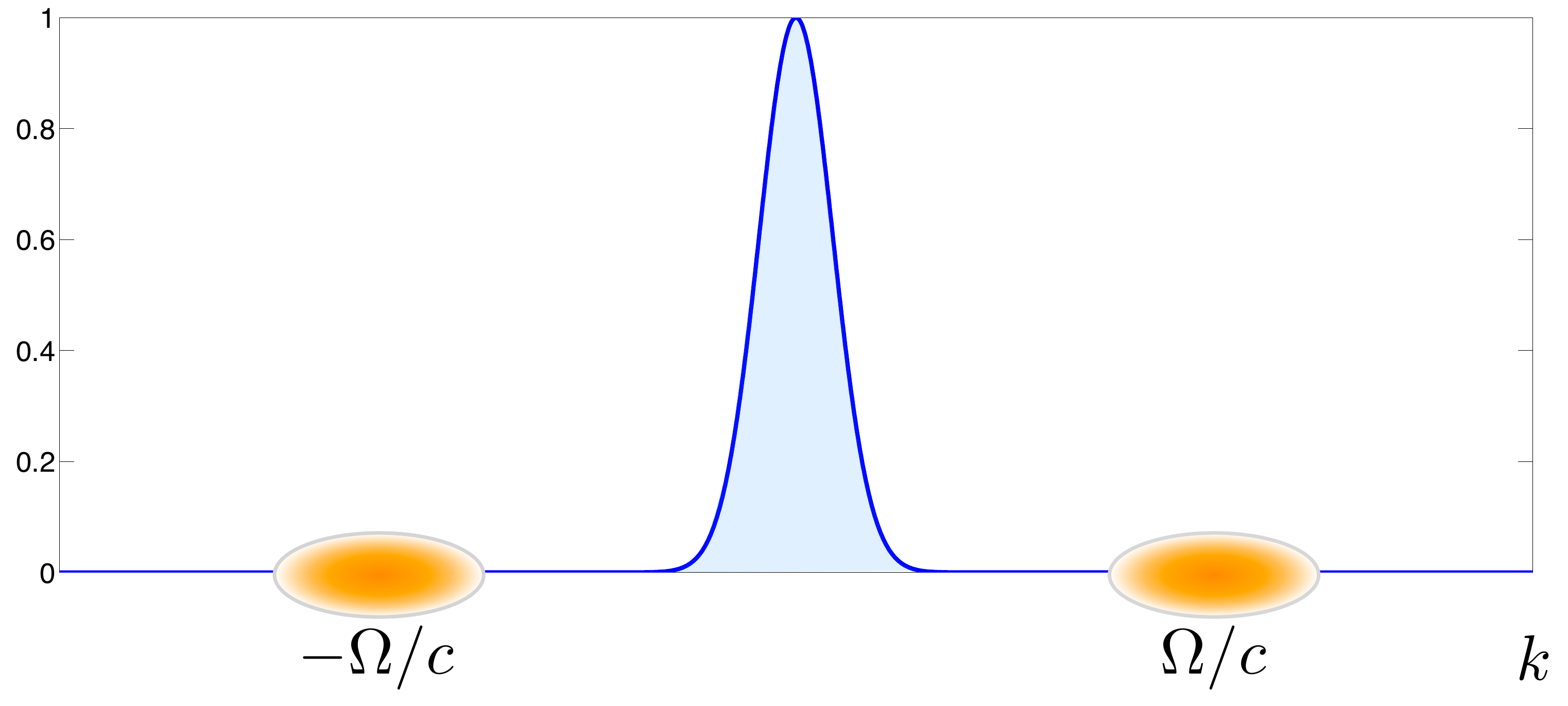}
\caption{A highly localised $\hat F(k)$ centred in 0 would practically suppress  the possibility of detection for the resonance frequencies to which the detector is most responsive, $k=\pm\Omega/c$. This results in a vanishing transition probability no matter what frequency we use to illuminate the detector.}
\label{fig1}
\end{figure}

So, as it is illustrated in fig. \ref{fig1}, if $\Omega\gg cL^{-1}$ the detector will not ever detect any signal even if it is a powerful pulse tuned to the transition frequency. Therefore, in order to be able to study relativistic settings, some modifications must be made to the model.

\begin{figure}[h]
\includegraphics[width=0.45\textwidth]{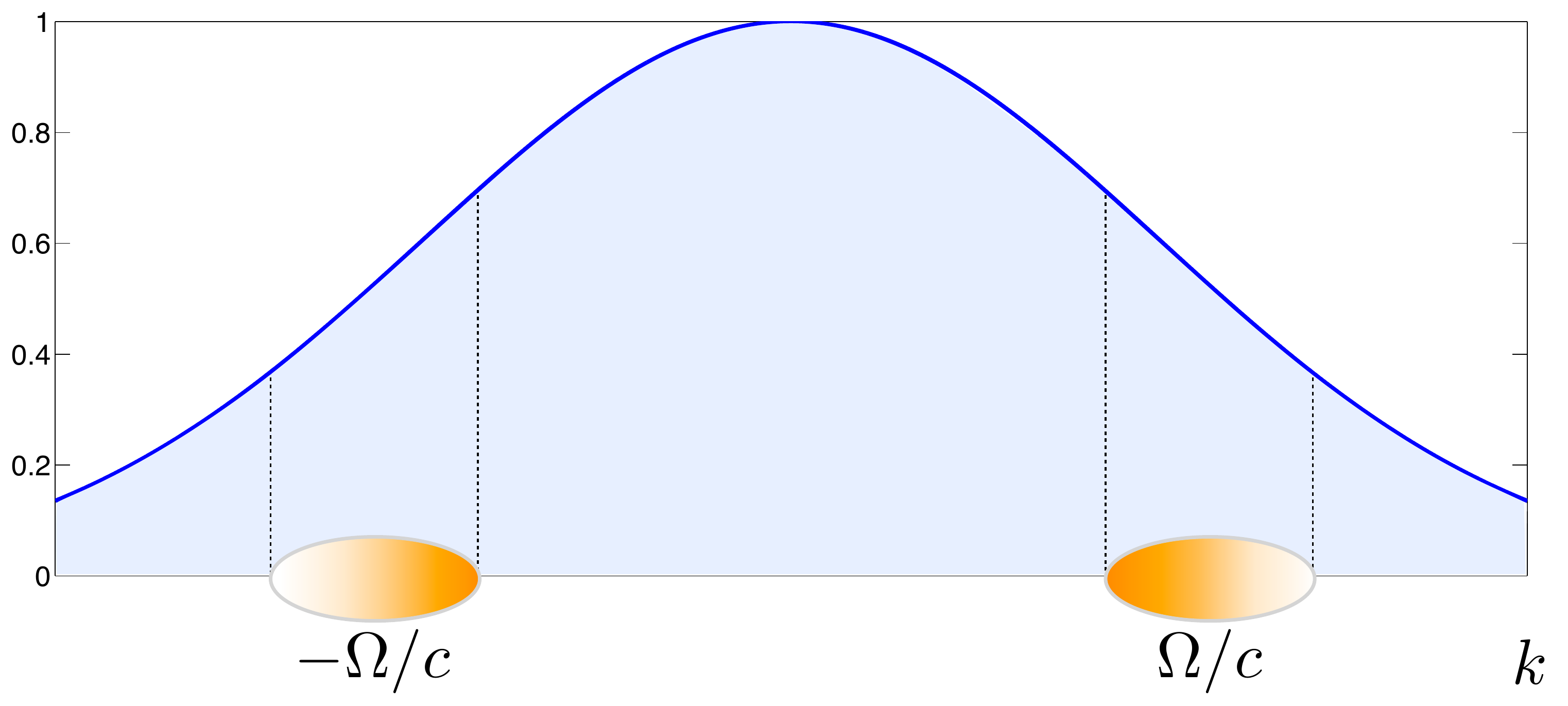}
\caption{A not-so localised $\hat F(k)$ centred in 0 would introduce an asymmetry in the detection of frequencies $\omega_k=\Omega\pm\delta$  $k=\pm\Omega/c$}
\label{fig2}
\end{figure}

One could argue that if the detector is very small with respect of the wavelength to which it is tuned (as it is the case of atoms), the Gaussian profile $\hat F(k)$ may cover the resonance regions. However, as seen in figure \ref{fig2}, if we analyze the probability of transition as a function of the frequency of the radiation with which the detector interacts, its spectral response will be asymmetric in the detuning between the detector natural frequency and the frequency of the radiation stimulating the transition $\Delta= \omega_k-\Omega$.

In other words, if the transition frequency is $\Omega$ and the radiation stimulating the transition is detuned from the energy gap of the detector by a small factor $\delta$,  the probability of transition will be positively weighted by $\hat F(k)$ if $\omega_k=\Omega-\delta$, and dampened if $\omega_k=\Omega+\delta$.

Although a similar asymmetry occurs in realistic atomic transitions (as detailed in section \ref{sec5}), the effect is so small that it can be neglected in most circumstances. In practice, no such effects are observed neither in atomic detectors nor in any other settings where quantum systems (like harmonic oscillators) are coupled to quantum fields.

When the size of the detectors increases as to become comparable with the wavelength to which they are tuned, e.g.  quantum microwave antennaee (for example flux or charge qubits in cQED), Rydberg atoms and cavity based detectors \cite{Cavities, Circuits}, the detector response is also symmetric in frequencies. Therefore the use of the Unruh-DeWitt detector presented above to model those scenarios (where the spatial profile is related to the natural dimension of the detector), can be problematic.

\section{Modulated oscillations in the spatial profile}\label{sec4}

In most realistic settings, the spectral response function of two level emitters is symmetric with respect to the resonance frequency, thus a small detuning should produce similar effects no matter if it is positive or negative. Also, as we discussed above, if the two level system size is comparable with the wavelength it is tuned to, the localized UdW model employed in the literature will dramatically fail to detect anything, even if it is the case of an intense pulse of radiation centred in the natural frequency of the detector's transition.

Taking these issues into account, we propose a modification of the way in which the UdW detector is spatially smeared. We will do so by feeding the spatial profile with information about the resonance frequency. For that matter, we will introduce a spatial profile which is strongly localized by a function $S(x)$, modulated by internal oscillations associated with the frequency the two level system is tuned to.

\begin{figure}[h]
\includegraphics[width=0.45\textwidth]{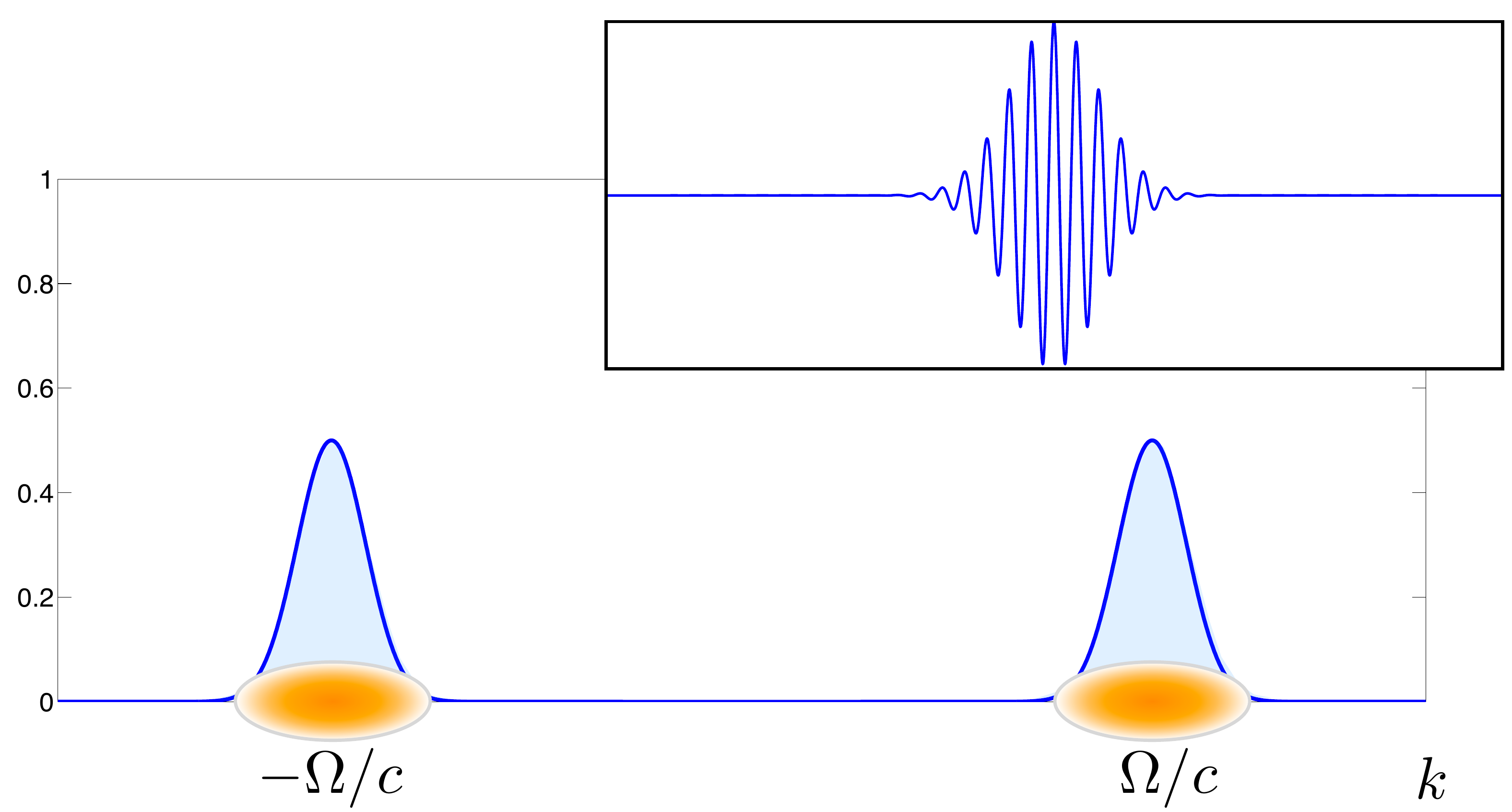}
\caption{A localised $\hat F(k)$ can be not centered in 0 by introducing a oscillating term in the spatial profile seen in the inset. The figure shows  symmetric detection zones centered in the frequencies $k=\pm\Omega/c$.}
\label{fig4}
\end{figure}

If the spatial profile is
\begin{equation}
F(x)=S(x)\cos\left(\frac{\Omega x}{c}\right)\label{trololo}
\end{equation}
then the spectral profile would be
\begin{equation}
\hat F(k)=\frac{1}{2}\left[\hat S(k-\Omega/c)+\hat S(k-\Omega/c)\right]
\end{equation}
which is a localised profile in frequencies around the two resonance regions. If we take $S(x)$ to be the Gaussian profile \eqref{gauss} then
\begin{equation}
\hat F(k)=\frac12\left(e^{\-\frac12 (k-\Omega /c)^2L^2}+e^{-\frac12 (k+\Omega /c)^2L^2}\right)
\end{equation}
which, as seen in figure \ref{fig4}, covers symmetrically the resonance regions.

By doing this we have the desired spectral response no matter the value of $\Omega$, and the detector is spatially localized around $x=0$ with a characteristic proper length $L$. We must stress that the introduction of the cosine factor in \eq{trololo} is intended only as a solution to the problem of the unphysical suppression of the transition rates. We do not claim that such a spatial profile is realised, for instance, in inertially moving two-level atoms (where the pointlike approximation is often valid and enough to produce physical results). However, this will not be the case when the UdW detector is used to model more exotic systems where the wavelength of the absorbed and emitted radiation is comparable with the size of the physical system. The problem of considering the physical form of $F(\bm{x})$ for regular atoms was tackled in section \ref{sec2}. 

Notice that we are not deriving this effective coupling from first principles. Rather, we are pointing out the limits of applicability of the UdW model to describe extended detectors when the wavelength of the radiation is comparable to their physical extension, and suggesting a way in which the phenomenology of such detectors can be effectively recovered. One can, however, understand this as a process of `antennization' (classical antennae, that are comparable with the wavelength of the radiation they are tuned to detect, have some periodical structure related to the wavelength they are resonant with). We are providing the extended detector with a spatial periodicity related to the radiation the detector is tuned to.

\section{Accelerated detectors}\label{sec5}

In order to provide a complete description of the localised detector model proposed in this note, in this section we will describe how to use this model to analyse arbitrary signals with a spatially smeared uniformly accelerated detector.

There is a well known problem with accelerating rigid bodies: the proper distance between two points of a solid accelerating with the same relativistic acceleration increases with time, eventually destroying the solid when the internal tension it supports is overrun by the relativistic effects.

The reasonable hypothesis for a physical detector is that it has to keep internal coherence. This  means that the internal forces that keep the detector together will prevent it from being further smeared due to relativistic effects up to some reasonable acceleration regimes. That means that, effectively, every point of the detector will accelerate with a different acceleration in order to keep up with the rest of its points. The natural formalism to treat this detector is the use of the well-known Fermi-Walker coordinates \cite{gravitation,Schlicht}.

Thus, the interaction Hamiltonian of a smeared uniformly accelerated rigid detector is
\begin{multline}
\label{Hamilera}
H_\text{I}(t)=g\!\!\int\!\! \frac{d k}{\sqrt{2\omega_k(2\pi)}}\!\! \int d \chi\ F( \chi)\left(\sigma^+e^{i\Omega \tau}+\sigma^-e^{-i\Omega \tau}\right) \\ \left(a^\dagger_{k} e^{i(\omega_k t( \chi,\tau) - kx( \chi,\tau))}+a_{k} e^{-i(\omega_k t( \chi,\tau)- kx( \chi,\tau))}\right)
\end{multline}
where $\bm \chi=(\chi,0,0)$ and $\tau$ are the Fermi-Walker coordinates associated with the trajectory of the detector.

These coordinates have the particularity that at every point on the trajectory $x(\tau)=(ct(\tau),x(\tau),0,0)$ the hyperplane which is orthogonal to the 4-velocity $u(\tau)=(c\dot{t}(\tau),\dot{x}(\tau),0,0)$ is the three-dimensional space which consists of all the events which are simultaneous to $x(\tau)$, where simultaneity is judged from the comoving inertial frame. We assume that we move only in one direction, so that $\chi_1=\chi,\chi_2=y=0,\chi_3=z=0$.

If we attach a dreibein to every such hyperplane
\begin{align}
e_{\chi_1}&=(c^{-1}\dot x(\tau),\dot t(\tau),0,0)\nonumber\\
e_{\chi_2}&=(0,0,1,0),\quad e_{\chi_3}=(0,0,0,1),
\end{align}
we can characterise every event $x_e$ in a neighborhood of the trajectory with $(\tau_e,\bm \chi_e)$.

These coordinates guarantee a rigid detector (where rigidity means that its 3-geometry as seen from its own momentary rest system is unchanged in the course of proper time). In contrast, in a Rindler frame (standard approach for pointlike detectors) every point of the detector accelerates with a different proper acceleration, so they cannot account for rigid detectors that have internal coherence. In the F-W frame the detector will accelerate coherently, so this models very well what would happen to an accelerated rigid-body.

The change of coordinates between the inertial system to the Fermi-Walker frame is given by
\begin{align}
\label{change}
\bm x(\tau,\bm \chi)=\bm x(\tau)+\chi^i \bm e_i (\tau),\quad t(\tau,\bm \chi)=t(\tau)+\frac{\chi^i e_i^0}{c}
\end{align}

For the uniformly accelerated observer, the trajectory (parametrised in terms of comoving time) is
\begin{equation}
x(\tau)=\left[\frac{c^2}{a}\sinh\left(\frac{a\tau}{c}\right),\frac{c^2}{a}\cosh\left(\frac{a\tau}{c}\right),0,0\right]
\end{equation}
The only relevant component of the dreibein is
\begin{equation}
e_{\chi_1}=\left[\sinh\left(\frac{a\tau}{c}\right),\cosh\left(\frac{a\tau}{c}\right),0,0\right]
\end{equation}
So, directly from \eqref{change} we read the change of coordinates
\begin{align}
\label{changec}
\nonumber t(\tau,\chi)&=\left(\frac{c}{a}+\frac{\chi_1}{c}\right)\sinh\left(\frac{a\tau}{c}\right)\\
\bm x(\tau,\chi)&=\left[\left(\frac{c^2}{a}+\chi_1\right)\cosh\left(\frac{a\tau}{c}\right),0,0\right]
\end{align}

Within this scheme we compute the probability of excitation of an accelerated detector responding to an arbitrary signal. In first order perturbation theory,
\begin{equation}
P=\left|g \right|^2\int_{\tau_0}^\tau d\tau' \int_{\tau_0}^\tau d\tau''\ e^{i\Omega (\tau'-\tau'')} \bra{y}\Psi(\tau'')\Psi(\tau')\ket{y}
\end{equation}
\begin{equation}\label{uno1}
\Psi(\tau)\!=\!\!\int\!\! \frac{F(\bm \chi)\,d {k}d {\chi}}{\sqrt{2c| k|(2\pi)}}\!\left(a_{\bm{k}} e^{i(\bm k\cdot \bm x(\bm \chi,\tau) -c|\bm k|  t(\bm \chi,\tau) )}\!+\!\text{H.c.}\right)\!
\end{equation}
where $\ket{y}$ is a general superposition of plane-wave field modes corresponding to a Minkowskian-shaped wavepacket, prepared in the lab, that we want to analyze with our detector,
\begin{equation}
\ket{y}= \left(\int d k \ y( k) a^\dagger_{k}\right).\ket{0}
\end{equation}

Let us evaluate the time-correlation function $W_y(\tau',\tau'')\equiv\bra{y}\Psi(\tau'')\Psi(\tau')\ket{y}$. The two $\chi$ integrals can be rewritten in terms of Fourier transforms greatly simplifying the expression of $W_y(\tau',\tau'')$. To do this we first note that
\begin{align}\label{uno2}
& k x( \chi,\tau) -ck  t( \chi,\tau) = L( k,\tau) \left(\chi+\frac{c^2}{a}\right)\nonumber\\
& L( k,\tau)=k e^{a\tau/c}
\end{align}

Considering that  $\omega=ck$, then the complex exponential argument depending on $\tau$ as taken directly from the amplitude in \eqref{uno1} and \eqref{uno2} goes as
\begin{equation}
\Omega \tau + \frac{c \omega}{a} e^{-a \tau/c}.
\end{equation}
So, taking derivatives, the condition for the stationary phase is as follows
\begin{equation}
\Omega  - \omega e^{-a \tau/c}=0.
\end{equation}

Now the condition is no longer time independent as in the inertial case \cite{Diegger}. Instead the resonance frequency $\omega_R$ will be 
\begin{equation}
\omega_R=\Omega e^{a\tau/c}.
\end{equation}
which is obviously the inertial resonance frequency but non-trivially Doppler-shifted due to the acceleration.

Now if we define $G^{\pm}( k,\tau)=\hat F[\pm L( k,\tau) ]$,
where $\hat F( k)$ is the Fourier transform of $F(\chi)$ as in \eqref{foutra}, we can rewrite $W_x(\tau',\tau'')=$
\begin{align}
&\!=\!\int \frac{\bar y(k)y(\kappa) d {k}d {\kappa}}{2(2\pi) c\sqrt{| k||\kappa|}}  G^+( k,\tau'')G^-( \kappa,\tau')e^{i\frac{c^2}{a}[L(\kappa,\tau')-L(  k,\tau'')]} ) \nonumber\\
&+\int \frac{|y(\kappa)|^2 d {k}d {\kappa}  }{2(2\pi) c| k|} G^+( k,\tau')G^-( k,\tau'')e^{i\frac{c^2}{a}[L(  k,\tau'')-L(  k,\tau')]} \nonumber\\
&+\int \frac{\bar y(k)y(\kappa)d {k}d {\kappa} }{2(2\pi) c\sqrt{| \kappa|| k|}} G^+(\kappa,\tau')G^-(     k,\tau'')e^{i\frac{c^2}{a}[L( k,\tau'')-L( \kappa,\tau')]}\nonumber
\end{align}
which can be further simplified if $F(k)=F(-k)$ (true for a Gaussian or Lorentzian profile), then we get  $G^+= G^-=G$ (although in general $G(k)\neq G(-k)$), and if the frequency profile of the signal $y(\omega)$ we want to analyse is chosen to be real,  we can rewrite $W_x(\tau',\tau'')=$
\begin{align}
&\!\!=\!\!\int\!\! \frac{y(k)y(\kappa)d {k}d {\kappa}}{(2\pi) c\sqrt{| k||\kappa|}}  G( k,\tau'')G( \kappa,\tau')\!\cos\!\Big[\frac{L(\kappa,\tau')\!\!-\!\!L(  k,\tau'')}{ac^{-2}}\Big] \nonumber\\
&+\int \frac{[y(\kappa)]^2d {k}d {\kappa}  }{2(2\pi) c| k|}  G( k,\tau')G( k,\tau'')e^{i\frac{c^2}{a}[L(  k,\tau'')-L(  k,\tau')]},
\end{align}
providing an operative expression for the response of a localized accelerated detector to a given signal.

\section{Conclusions}\label{conc}

In this work, we have analysed the problem of wavepacket detection by an UdW model. 

By appealing to phenomenological considerations, we have argued that in scenarios where our detector has to respond to a given frequency, the spatial profile considered must verify certain properties. In particular, we have studied the origin of such a profile function for the case of an atomic detector by takink the task of deriving a UdW equation from first principles, relating the smeared UdW model to the usual $\bm{p}\cdot\bm{A}$ form of the QED interaction coupling atoms to the electromagnetic field. We have shown what differences between the models actually result from this calculation. As an outcome, we have shown a way of relating the smearing profile used in the UdW case with the electronic wavefunction of the relevant orbitals of an atom.

Going beyond this atomic example, and especially, when considering the case of detectors comparable with the wavelength to which they are tuned, we show that some information about the spectral response of the detector must be fed in general to the spatial profile. Otherwise the detector will not have the expected behaviour and will dramatically fail to detect radiation on resonance with the two-level system transition. 

To solve these problems, we suggest to introduce a spatial oscillation of the profile, which will make the detector tune to the resonance frequency regardless of its size and configuration.

Not all the spatial profiles for the UdW model would be compatible with the experimental response of accelerated particle detectors: the existence of some monopole (or dipolar) momentum that couples the atom to the field with a given characteristic transition frequency requires those oscillations introduced in the spatial profile to reproduce spectra centred in the characteristic transition frequency of the detector . If one thinks of that profile as being something like a charge distribution, then those oscillations would be the responsible for the appearance of the momentum that correctly couples it to the field.

Completing our proposal, we have explained how to use this formalism while calculating the probability of detection of a wavepacket for an accelerated detector. 

Finally note that, in parallel with this work, an analysis of the transition rates of smeared UdW detectors coupled to different kinds of physical field modes and undergoing different relativistic motion is being carried out by Lee and Fuentes \cite{Anta}.

\section{Acknowledgements}

We thank Achim Kempf, Markus M\"ulle and Antony Lee for interesting discussions during the RQI-N 2012 conference held at Perimeter Institute. We are very grateful to Jorma Louko for his insightful comments and feedback and to Juan Le\'on and Borja Peropadre for fruitful discussions. Research at Perimeter Institute is supported by the Government of Canada through Industry Canada and by the Province of Ontario through the Ministry of Research \& Innovation. E. M-M. acknowledges support of the Banting Fellowship programme. This work was also supported by the Spanish MICINN Project FIS2011-29287 and CAM research consortium QUITEMAD S2009-ESP-1594. M. del Rey was also supported by a CSIC JAE-PREDOC grant and Fundaci\'on Bot\'in.

\end{document}